\documentclass[12pt]{article}

\usepackage{amsmath,amssymb,graphicx} 
\usepackage{epsf}
\usepackage{cite}

\def\beq{\begin{equation}}
\def\bea{\begin{eqnarray}}
\def\eea{\end{eqnarray}}
\def\de{\partial}

\newcommand{\sfrac}[2]{{\textstyle\frac{#1}{#2}}}

\def\centeron#1#2{{\setbox0=\hbox{#1}\setbox1=\hbox{#2}\ifdim
\wd1>\wd0\kern.5\wd1\kern-.5\wd0\fi
\copy0\kern-.5\wd0\kern-.5\wd1\copy1\ifdim\wd0>\wd1
\kern.5\wd0\kern-.5\wd1\fi}}
\def\ltap{\;\centeron{\raise.35ex\hbox{$<$}}{\lower.65ex\hbox{$\sim$}}\;}
\def\gtap{\;\centeron{\raise.35ex\hbox{$>$}}{\lower.65ex\hbox{$\sim$}}\;}

\def\singlespaced{\baselineskip=\normalbaselineskip}

\def\dslash{\not{\hbox{\kern-2pt $\partial$}}}
\def\Dslash{\not{\hbox{\kern-4pt $D$}}}
\def\Oslash{\not{\hbox{\kern-4pt $O$}}}
\def\Qslash{\not{\hbox{\kern-4pt $Q$}}}
\def\pslash{\not{\hbox{\kern-2.3pt $p$}}}
\def\kslash{\not{\hbox{\kern-2.3pt $k$}}}
\def\lslash{\not{\hbox{\kern-2.3pt $l$}}}
\def\qslash{\not{\hbox{\kern-2.3pt $q$}}}
\def\epsilonslash{\not{\hbox{\kern-2.3pt $\epsilon$}}}

\newcommand{\newc}{\newcommand}
\newc{\qbar}{{\overline q}}
\newc{\Kahler}{K\"ahler }
\newc{\deltaGS}{\delta_{\rm GS}}

\begin{document}
\begin{titlepage}
\begin{flushright}
{\tt hep-th/0312117}\\
\end{flushright}


\begin{center}
{\huge \bf Massive Gravity on a Brane }
\end{center}


\begin{center}
{\bf Z. Chacko$^{a}$,
 M.L. Graesser$^{b}$,
 C. Grojean$^{c,d}$ and L. Pilo$^{e}$}
\end{center}

\begin{center}
$^{a}${\it Department of Physics, University of California at Berkeley and \\
Lawrence Berkeley National Laboratory, Berkeley, CA 94720, USA} \\
\vspace*{0.1cm}
$^{b}$ {\it California Institute of Technology, 452-48, Pasadena, CA 91125} \\
\vspace*{0.1cm}
$^{c}$   {\it Service de Physique Th\'eorique,
CEA Saclay, F91191 Gif--sur--Yvette, France} \\
\vspace*{0.1cm}
$^{d}$ {\it Michigan Center for Theoretical Physics,
Ann Arbor, MI 48109, USA}\\
\vspace*{0.1cm}
$^{e}$ {\it Dipartimento di Fisica `G. Galilei', Universit\`a di Padova\\
INFN, Sezione di Padova, Via Marzolo 8, I-35131 Padua, Italy }\\
\vspace*{0.3cm}
{\tt zchacko@thsrv.lbl.gov,  graesser@theory.caltech.edu,}\\
{\tt grojean@spht.saclay.cea.fr, pilo@pd.infn.it}
\end{center}


\begin{abstract}
\noindent
At present no theory of a massive graviton is known that is consistent with experiments at both long
and short distances. The problem is that consistency with long distance experiments requires  the
graviton mass to be very small. Such a small graviton mass however implies an ultraviolet cutoff for
the theory at length scales far larger than the millimeter scale at which gravity has already been
measured. In this paper we attempt to construct a model which avoids this problem. We consider a
brane world setup in warped AdS spacetime and we 
investigate the consequences of writing a mass 
term for the graviton on an infrared brane  where
the local cutoff is of order a large (galactic) distance scale. The advantage of this setup is that the
low cutoff for physics on the infrared brane does not significantly affect the predictivity of the
theory for observers localized on the ultraviolet brane. For such observers the predictions of this
theory agree with general relativity at distances smaller than the infrared scale but go over to those
of a theory of massive gravity at longer distances. A careful analysis of the graviton two-point
function, however, reveals the presence of a ghost in the low energy spectrum. A mode decomposition of
the higher dimensional theory reveals that the ghost corresponds to the radion field. We also
investigate the theory with a brane localized mass for the graviton on the ultraviolet brane, and show
that the physics of this case is similar to that of a conventional four dimensional theory with a
massive graviton, but with one important difference: when
the infrared brane  decouples and  the would-be massive
graviton gets heavier than the
regular Kaluza--Klein modes, it becomes unstable and it has a
finite width to decay off the brane into the continuum of Kaluza--Klein states.

\end{abstract}

\end{titlepage}

\newpage

\setcounter{footnote}{0} \setcounter{page}{2}
\setcounter{section}{0} \setcounter{subsection}{0}
\setcounter{subsubsection}{0}


\section{Introduction}

Recently there has been considerable interest in theories of gravitation which deviate from Einstein's
gravity at very long distances (for example,
\cite{KMP,GRS,CEH,DGP,GG}). However there has been no
consistent theory yet proposed which is consistent with all observations at both macroscopic and
microscopic length scales \cite{PRZ,AGS,DKP,LPR,Rub,DL}. Conceptually
perhaps the simplest modification is 
a mass term for the graviton {\cite{fierzpauli}}.
However this theory suffers from three 
difficulties which are typical of theories of massive gravity as a
whole. 
Firstly, unless the mass term has the Fierz-Pauli form the 
theory has a ghost \cite{fierzpauli,VV}. 
Secondly, 
for a graviton mass $m_g$ this theory 
has a cutoff $(m_g^4 M_{4})^{\frac{1}{5}}$
{\cite{AGS}}, where $M_4$ is the Planck scale. This cutoff 
is much too low for the theory to be
simultaneously consistent with experiments at 
microscopic and macroscopic scales. Thirdly, even for
arbitrarily small graviton mass the longitudinal component of the massive graviton does not decouple
from sources. This fact, which was first observed by van Dam, Veltman and Zakharov {\cite{VV}}, implies
that the tensor structure of the gravitational interaction deviates from that of Einstein gravity.
While there are indications that suitable ultraviolet completions may be free of the latter problem
{\cite{V,DGP2,Gr,Po}} to date no completely satisfactory candidate theories
are known {\cite{DKP,LPR,Rub,DL}}.

In the absence of a known Higgs mechanism for gravity it might seem that these problems pose an
insurmountable obstacle in constructing any experimentally viable theory of a massive graviton. However
a closer examination suggests that this need not be the case. Consider the five dimensional 
brane model of Randall and Sundrum (RS) {\cite{RS2}}. This is a simple example of a theory where the
local cutoff varies from point to point in the higher dimensional space. In particular in the far
infrared the cutoff of the theory is below the millimeter scale, where gravity has been measured in the
laboratory. This low cutoff is completely consistent with these experiments because physics
measurements on the brane at any four momentum scale $p$ are exponentially insensitive to points in the
bulk where the local cutoff is lower than the scale $p$.

The success of this theory suggests a means whereby the problems normally associated with theories of
massive gravity can be avoided. To the single brane model of Randall and Sundrum we add a second brane
deep in the infrared such that the compactification radius, which is the inverse mass of the lightest
Kaluza--Klein state, is of order galactic size. On this second brane we add a brane localized
mass term for the graviton. The extra dimension just corresponds to an interval with two boundaries (see~Ref.~\cite{NoHiggs} for similar constructions for gauge theories and their application
for the problem of electroweak symmetry breaking).
Since two point correlators with external legs on the ultraviolet brane are exponentially insensitive
to physics on the infrared brane for four momenta above the compactification scale we expect that
conventional Einstein gravity will be reproduced on the ultraviolet brane at distances shorter than the
compactification scale. However, 
at distances longer than the compactification scale, 
the theory is
sensitive to infrared physics, with the consequence that 
below this scale 
the theory with a Fierz-Pauli mass term 
is expected to
resemble the four-dimensional Fierz--Pauli 
theory of a massive graviton. This then would be a concrete
realization of a an experimentally viable theory of massive gravity.  In this paper we investigate this
proposal in detail. This picture is correct, at the price of a serious drawback though:
while the theory does indeed
reproduce Einstein gravity at sub-galactic length scales the low energy spectrum in the four
dimensional effective theory contains, in addition to a massive graviton, a ghost state. A mode
decomposition of the higher dimensional theory reveals that it is the radion field which is a ghost.
{\footnote{Under certain circumstances theories with ghosts may in fact be
viable~\cite{Nima,CJM}. However we do not pursue this possibility here.}}
We also find that this conclusion is rather general: even allowing 
for a non-Fierz-Pauli mass term on the IR brane the radion field 
is always a ghost. The analysis of the more general case is 
provided in Appendix B. 

In the following sections we explore in detail the model we are investigating. We compute the graviton
two point function with external legs on the ultraviolet brane and show that while the predictions of
the theory agree with those of general relativity for observers on the ultraviolet brane probing
distance scales shorter than the compactification radius, the light states consist of a massive
graviton and a ghost. We then perform a mode decomposition of the linearized theory for both the
transverse traceless modes and the radion. This reveals that the transverse traceless modes of the
theory without a mass term smoothly go over to the transverse traceless modes of the theory with a mass
term as the mass term is turned on. However the same is not true of the radion. Instead, the radion
changes discontinuously into a ghost as soon as the mass term is turned on.
 We also study the theory with a mass term on
the ultraviolet brane and show that the predictions of this theory for observers on the ultraviolet
brane agree with those of a four dimensional theory with a massive graviton.
However there is one
important difference: if the extra dimension is sufficiently large that the would-be graviton is
heavier than the lightest regular Kaluza--Klein states, then it becomes unstable and  it has a finite though small width to decay into the regular Kaluza--Klein states.

\section{Brane Localized Fierz--Pauli Mass Term}
\label{sec:EinsteinEqs}

\subsection{Bulk equations of motion}

We consider a brane world model whose dynamics is governed by the following action:\footnote{Our
conventions correspond to a mostly plus signature $(-+\ldots+)$ and the definition
of the curvature is such that an Euclidean sphere has a positive curvature. Bulk coordinates
will be denoted by capital Latin indices and brane coordinates by Greek indices.}
\begin{equation}
\mathcal{S}
=
\int d^5x \sqrt{|g|}
\left(
\frac{\mathcal{R}}{2\kappa^2_5} - \Lambda
+ (-\sigma_i + \mathcal{L}_i) \, \delta (\sqrt{g_{55}} (z-z_i))
\right)
\end{equation}
$z_{i=1,2}$ are the locations of the two branes, $\Lambda$ is the bulk cosmological constant, and
$\kappa_5$ is related to the 5D Planck (fundamental) scale $M_5$ by $M_5^3 = 1/2 \kappa^2_5$,
$\sigma_i$ are the brane cosmological constants (tensions) and  $\mathcal{L}_i$  are
Lagrangian densities describing some boundary localized matter fields.
We will fine-tune the bulk and brane cosmological constants such that the background geometry
corresponds to the well-known  Randall--Sundrum solution:
\begin{equation}
ds^2 = \left({R \over z} \right)^2
\left(\eta_{\mu \nu} dx^\mu dx^\nu + dz^2 \right)
\end{equation}
with $R^{-1}=\sqrt{-\kappa_5^2 \Lambda/6}$, $\sigma_{UV}= 6/(\kappa_5^{2}R)$ and
$\sigma_{IR}= -\sigma_{UV}$. The location of the branes are
such that the warp factor, $R/z$, is set to one on the ultraviolet (UV) brane ($z_1=R$),
and it is  exponentially smaller on the infrared (IR) brane ($z_2=R^\prime \gg R$).

The aim of this paper is to study the spectrum of the physical excitations
when non-trivial gravitational interactions are introduced on the branes.
We thus need to consider  gravitational fluctuations around the RS background solution:
\begin{equation}
ds^2 = e^{2A} \, \left( \eta_{{}_{MN}} + h_{{}_{MN}}\right) d x ^{{}_M} d x^{{}_N} \, ,
\quad  \mathrm{with} \quad A=  -\ln (z/R)  .
\end{equation}

In the bulk, the Einstein's equations are of course independent of
brane interaction terms. At the linear level and in absence of any matter beside
the bulk cosmological constant, these equations read
\begin{equation}
E^{(1)}_{MN} \equiv G_{MN}^{(1)} + \kappa_5^2 \Lambda \,e^{2A} h_{MN} = 0,
\end{equation}
where $G_{MN}^{(1)}$ is the linear piece of the Einstein tensor.
Using the Einstein equations of the background solution, we finally arrive at
\begin{eqnarray}
\label{lineareeq1}
E^{(1)}_{\mu \nu}
&=&
\sfrac{1}{2} \left(
\partial_\mu \partial^\sigma h_{\nu \sigma} + \partial_\nu \partial^\sigma h_{\mu \sigma}
- \Box h_{\mu \nu} - \partial_{\mu} \partial_{\nu} h
\right)
- \sfrac{1}{2} \left(
\partial_\sigma \partial_\rho h^{\sigma\rho} - \Box h
\right) \eta_{\mu\nu}
\nonumber \\
&&
- \sfrac{1}{2} (h^{\prime\prime}_{\mu \nu}-  h^{\prime\prime} \eta_{\mu \nu})
- \sfrac{3}{2} (h^{\prime}_{\mu \nu}-  h^{\prime} \eta_{\mu \nu}) A^\prime
\nonumber \\
& &
- \sfrac{1}{2} \partial_\mu \partial_\nu h_{55}
+ \sfrac{1}{2}  (
\Box h_{55}
- 3 A^{\prime\prime} h_{55}
- 3 A^\prime h_{55}^\prime
- 9 {A^\prime}^2   h_{55}
)  \eta_{\mu \nu}
\nonumber\\
& &
+\sfrac{1}{2} \left(
\partial_\mu h_{\nu 5}^\prime
+ \partial_\nu h_{\mu 5}^\prime
\right)
- \partial^\sigma h_{\sigma 5}^\prime \,
\eta_{\mu \nu}
+ \sfrac{3}{2}
A^\prime \left( \partial_\mu h_{\nu 5} +   \partial_\nu h_{\mu 5} \right)
-3 A^\prime \partial^\sigma h_{\sigma 5}\,  \eta_{\mu\nu},
 \\
\label{lineareeq2}
E^{(1)}_{\mu 5}
&=&
\sfrac{1}{2} \left(\partial^{\sigma} h_{\sigma \mu} - \partial _{\mu} h \right)^{\prime}
+ \sfrac{3}{2} A^{\prime} \partial _{\mu} h_{55}
+ \sfrac{1}{2} \partial_\mu \partial^\sigma h_{\sigma 5}
- \sfrac{1}{2} \Box  h_{\mu 5}
+ 3 A^{\prime \prime} h_{\mu 5}
- 3 {A'}^2 h_{\mu 5}, \\
\label{lineareeq3}
E^{(1)}_{55}
&=&
- \sfrac{1}{2} \left( \partial^{\sigma} \partial^{\rho}
h_{\sigma \rho} - \Box h \right)
+ \sfrac{3}{ 2} A^{\prime } h^{\prime}
-6 {A'}^2 h_{55}
-3 A^\prime \partial^\sigma h_{\sigma 5},
\end{eqnarray}
with the following conventions: 4D indices are raised and lowered using the flat Minkowski
metric, $h$ is the 4D trace $ {h_{\mu}{}}^{\mu}$, $\Box = \partial^\sigma \partial_\sigma$
and a prime denotes a derivative with respect to the $z$ coordinate.

The bulk equations are obviously covariant under an infinitesimal
general coordinate transformation that reads at the linear order:
\begin{eqnarray}
&
\delta x^{{}_M} =  \xi^{{}_M} ,
\\
          \label{eq:deltah}
&
\delta {h}_{\mu \nu} =
- \partial_\mu \xi_{\nu} -  \partial_\nu \xi_{\mu}
-2 A^{\prime} \xi^5 \, \eta_{\mu \nu },
\
\delta {h}_{\mu 5}  = - \xi_{\mu}^\prime  - \partial_\mu \xi^5,
\
\delta {h}_{55}  =   - 2 \left(\xi^5 e^A \right)^{\prime} e^{-A} .
\label{generaldeltah}
\end{eqnarray}
Clearly, this reparametrization invariance allows to restrict ourselves to
generalized  Gaussian normal (GGN) systems of coordinates:
\begin{equation}
	\label{GGN}
h_{55}=h_{\mu 5}=0 \, , \qquad  \text{brane embeddings: } z=f_i(x), \; i=1,2 \;.
\end{equation}
Within these generalized Gaussian normal gauges, there is still a residual
reparametrization invariance involving  arbitrary functions, $\zeta$ and $\zeta^\mu$, of the 4D
coordinates :
\begin{eqnarray}
	\label{eq:redGGN5}
&
\xi^5(x,z)
=
{z \over R} \,  \zeta (x),
\\
	\label{eq:redGGNmu}
&
\xi_{\mu}(x,z)
=
\zeta_\mu (x) - \sfrac{1}{2}  {z^2 \over R} \, \partial_\mu \zeta (x),
\end{eqnarray}
and  the transformation of the metric fluctuations is
\begin{equation}
\delta {h}_{\mu \nu} =
{z^2 \over R} \, \partial_\mu\partial_\nu \zeta
+{2 \over R} \,   \zeta\, \eta_{\mu \nu}
-  \partial_{\mu} \zeta_{\nu}  - \partial_{\mu} \zeta_{\nu}.
\end{equation}
Clearly, with an appropriate choice of $\zeta$ we can
maintain the GGN gauge fixing conditions~(\ref{GGN}) and
straighten one of the branes which will now be located at a constant
$z$. This gauge choice for which the
UV (IR) brane is straight will be called
GNUV (GNIR),  generalized Gaussian normal gauge with respect to the UV
(IR) brane. There is finally a third special generalized Gaussian normal gauge
for which the 4D fluctuations are TT, ie, traceless, $h=0$, and transverse,
$\partial^\sigma h_{\sigma\mu}=0$.
\footnote{To see this, first note that the residual
gauge invariance can be used to set $h=0$ and $\partial^{\nu}
h_{\nu \mu}=0$ at a point $z=z_0$. Then the $(\mu 5)$ equation
implies that $\partial_{\mu} h - \partial^{\nu}
h_{\nu \mu}=0$ everywhere in the bulk. Using this result the $(55)$
equation then implies $h^{\prime}=0$ everywhere. But since
$h=0$ at a point, it is zero everywhere. Then
$\partial^{\nu} h_{\mu \nu}=0$ in the bulk.} This gauge
will be denoted GNTT.
Each gauge has its own advantage: in
the GNTT gauge it will be easy to solve the
bulk equations of motion while in the GNUV and GNIR gauges it will be
easy to solve
the boundary conditions.
In the following sections we will explain how these
different gauges are related to each other
depending on the interactions and the matter localized on the
branes.  Finally note that the GNUV and GNIR gauges
still possess
usual 4D reparametrization invariance
associated to the $\zeta^\mu(x)$.

\subsection{Boundary conditions in presence of brane mass term}

We now
want to add some brane localized interactions for the
gravitational degrees of freedom. More precisely,
we are interested in a localized
mass term. Working in the generalized
Gaussian system of coordinates in which the brane
we want to add the mass term on is straight,
the mass term is for simplicity chosen to be of the Fierz--Pauli 
form~\footnote{The powers of the warp factor are determined by the
requirement that in the coordinate
system where $e^A=1$ at the
IR  brane say, the boundary condition on that brane
is independent of the warp factor on the UV brane.}:
\begin{equation}
\label{fpbrane}
{\cal L} = - \sfrac{1}{8} \, f_{UV}^4
\int d^4 x  \, ({h}_{\mu \nu} h^{\mu \nu} -
{h}^2 )_{|z=R}
-\sfrac{1}{8} \, f_{IR}^4 \frac{R^4}{R^{\prime\,4}}
\int d^4 x  \, ( h_{\mu \nu} h^{\mu \nu} - h^2 )_{|z=R^\prime}
.
\end{equation}
Of course, since the mass terms~(\ref{fpbrane})  explicitly break
general coordinate invariance, their forms
will not be the same in different system of coordinates
and it will have to be determined
by coordinate transformation from the appropriate GGN gauge.
\footnote{In an abuse of language we will still refer
to generalized coordinate transformations
as ``gauge transformations'',
even though this symmetry
is
explicitly broken.} 
Note in particular that the two mass terms are not written in the same coordinate systems: the UV mass term is written in the GNUV gauge while the IR mass term is written in the GNIR gauge.

More general, non-Fierz--Pauli mass terms 
may also be considered. For brevity, the analysis 
of these more general mass terms is provided 
in the Appendix B. The conclusions of this 
and subsequent sections is unchanged in 
the more general case : an additional state is 
present, but it decouples and the radion is always a 
ghost.

The effect of the mass term and additional matter localized
on the brane is to modify
the usual Neumann boundary condition for the metric fluctuations.
The new boundary condition
gets simplified on the GGN gauge where the brane
is straight.
In the GNUV gauge, the boundary condition at the UV brane
is
\begin{equation}
h_{\mu\nu}^\prime = -\kappa_5^2 (S_{\mu\nu}^{UV}
- \sfrac{1}{3} \eta _{\mu \nu} S^{UV} - f^4_{UV} h_{\mu\nu}),
\end{equation}
where $S_{\mu\nu}^{UV}$ is the stress-energy tensor for the
matter localized on the UV brane,
$S^{UV}$ is its trace, $S_{\mu\nu}^{UV} \eta^{\mu\nu}$.
Similarly the boundary condition at the IR brane is
\begin{equation}
h_{\mu\nu}^\prime = \kappa_5^2 (S_{\mu\nu}^{IR} - \sfrac{1}{3}
\eta_{\mu \nu} S^{IR}) - \kappa^2_5
f^4_{IR} \frac{R}{R^\prime} \, h_{\mu\nu}  .
\end{equation}
where warp factors have been absorbed into our definition
of $S^{IR}_{\mu \nu}$.
These boundary conditions
are obtained by varying the brane localised mass term
action given above, and then adding that to the left-side of the linearized
Einstein equations.
The boundary conditions are then obtained by requiring the cancellation of the boundary
terms in the variation of the action (see for instance~\cite{NoHiggs} for an analogous computation in gauge theories).  For later convenience it is
useful to introduce the following parameters that have the dimension of a mass
\begin{equation}
\lambda_{IR(UV)} = \kappa^2_5 f^4 _{IR(UV)}~.
\end{equation}

We can find the graviton Kaluza--Klein spectrum by solving the bulk
equations for
transverse and traceless excitations,
supplemented by the above boundary consitions in the absence of matter on the branes.
The presence of the brane mass terms lifts  the zero mode from the
spectrum. For instance
in the presence of a mass term on the IR brane only and in the limit
$\lambda_{IR} \ll R^{-1}$, the lightest spin-2 state has a
mass given by (see Section 4.1 for details)
\begin{equation}
m_0^2 \sim 2 \, \frac{\kappa_5^2  f^4_{IR}}{R} \left( \frac{R}{R^\prime} \right)^4
\sim {2 \over M^2_{Pl}} \left( {R \over R^{\prime}} f_{IR} \right)^4
\end{equation}
where $M^2_{Pl} \sim R/\kappa^2_5 $ is approximately the
four-dimensional Planck mass.

The spectrum of light states with mass below the
effective compactification scale $1/R^{\prime}$
is seen to also contain
a massless scalar, the radion.
Part of this paper is devoted to
the identification of the properties of this perturbation.

\subsection{State counting}

In this section we count the number of degrees of freedom in the gravity theory with a brane localized
Fierz--Pauli mass term. In the theory without such a mass term the spectrum consists of a massless
spin-2 field with two polarizations, a radion and a tower of massive spin-2 Kaluza--Klein resonances
with five polarizations each. Here we show that once the mass term is introduced the spectrum changes
only in that the lightest spin-2 field is now massive and therefore has the five polarizations
associated with a massive spin-2 particle.

In doing so we will make use of the important result that in the presence of the brane mass term and in
the absence of any additional matter on the {\it massive} brane, the gauge GNIR(GNUV) is equivalent to
the GNTT gauge. In subsequent sections we will also make use of this result.

Before demonstrating this, we note that this is
similar to the situation with a massive
$U(1)$ vector boson $A_{\mu}$. There one has no gauge
invariance, but a priori
four degrees of freedom. However, the equations of
motion for the vector boson imply that $\partial_{\mu} A^{\mu}=0$,
eliminating one of the unphysical perturbations. Thus the
theory describes three fluctuating degrees of freedom,
the correct number for a massive spin--1 particle.

Similarly, a massive graviton in four dimensions
a priori describes ten degrees of freedom,
but has no gauge invariance. As with the
massive vector boson, one finds that for
the Fierz--Pauli mass term, the equations
of motion imply that the metric is transverse and
traceless. This eliminates five perturbations
leaving five, which is the correct number
for a massive spin-2 particle~\cite{fierzpauli,boulwaredeser}.

The situation with the brane localized
mass term is similar, but naively worse.
This is because we are describing a
five dimensional gravitational theory,
which a priori has fifteen degrees of
freedom.
Since the brane mass term explicitly breaks
general coordinate invariance,
there is a concern that
additional states which were previously eliminated
by the gauge invariance are now reintroduced.
This would be a disaster for the model, just as for a
massive graviton theory with a mass-term
in a non-Fierz--Pauli combination.

However,
just as in the massive vector and graviton examples
given above, we shall see that
the equations of motion imply
that for a Fierz--Pauli mass term
defined in GNIR(GNUV) gauge, the metric is
additionally TT in absence of any additional matter on the brane. Thus the only degrees of
freedom are massive gravitons which involve only five
physical polarizations  and a massless
radion associated with the movement of the
brane. In addition, since this result
will follow from the properties of the bulk
equations of motion and the IR(UV) boundary
condition, these conclusions are unchanged
if a source is placed on the opposite UV(IR) brane.

To see this, let us consider the case of a mass term added on the IR brane and let us work in
the GNIR gauge where the IR brane is straight.
The IR boundary
condition for the combination
$H_{\mu} \equiv \partial^{\nu} h_{\nu \mu}-\partial _{\mu}
h$ is
\begin{equation}
\partial _z {H_{\mu}}_{|z=R^{\prime}}
=
- \lambda_{IR} {R \over R^{\prime}}  {H_{\mu}}_{|z=R^{\prime}} ~.
\end{equation}
Now the $(\mu 5)$ equation implies
that $\partial_z H_{\mu}=0$,
so the left-side
of the equation above vanishes identically. Since
$\lambda_{IR} \neq 0$, we learn that $H_{\mu}$
vanishes
at the location of the IR brane.
But since by the $(\mu 5)$
equation $H_{\mu}$ is constant in the bulk,
we find that it actually
vanishes identically. Using this result
the $(55)$ component of the bulk Einstein equations implies
that $\partial_z h =0$ identically.
Next consider the IR boundary condition
again, but written as $\partial _z {h_{\mu \nu}}_{|z=R^{\prime}}=
-\lambda_{IR}  R/R^{\prime} \, {h_{\mu \nu}}_{|z=R^{\prime}}$.
The trace of this implies $\partial _z h_{|z=R^{\prime}}
= - \lambda_{IR} R/R^{\prime} \, h_{|z=R^{\prime}} $,
which when combined with the previous result
implies that $h_{|z=R^{\prime}}=0$.
But $\partial_z h=0$, so $h$~vanishes in the
bulk.
From
this result and $H_{\mu}=0$,
it follows that $\partial ^{\nu} h_{\nu \mu}=0$
identically.

The introduction
of a brane localized Fierz--Pauli mass term (\ref{fpbrane})
therefore implies that the metric is transverse and traceless in the GGN gauge where the brane is straight (the GNIR gauge is also GNTT).
Thus at the massive level there are only
five degrees of freedom, corresponding to the
helicity states of a massive graviton.
The brane localized Fierz--Pauli mass
term does not introduce any additional massive
degrees of freedom that were not already present
in the RS model.

These results may be understood
by noting that
the model
still has a large residual general invariance generated by
coordinate transformations that
vanish at the location of the brane.
Referring to
(\ref{generaldeltah}), this requirement implies
 ${\xi ^{\mu}}_{|z=R^{\prime}}=0$ and
${\xi^5}_{|z=R^{\prime}}=0$.
Thus the only gauge transformations explicitly broken
by the brane mass term are those  associated with
the would--be
zero mode graviton and the bending of the brane.

\section{Two-Point Function Analysis}

In this section we obtain expressions for the graviton two-point correlator with external legs on the
UV brane.{\footnote{The analogous calculation for the case of a gauge field
with a brane localized mass term may be found in {\cite{CP}}.} We consider first the case
with a mass term on the UV brane and then the case with a mass term on
the IR brane. This calculation serves two purposes. We will be able to determine the extent to which
observers on the UV brane find the theory to deviate from Einstein's gravity at any particular length
scales. We will also be able to determine the masses of the light modes in the four dimensional
effective theory. This is precisely the physics we are most interested in determining.

The metric
perturbation in the linear approximation
created by a source $S$ will be given by
\begin{equation}
h_{MN}(x,z) = \int d^4 x^{\prime} dz^\prime
\sqrt{|g|} \, \Delta^{PQ} _{MN}(X,X^{\prime}) S_{PQ}(x^{\prime},z^\prime) ~,
\end{equation}
where $\Delta$ is the Green's function.
We are mostly interested in physics for an observer on the Planck brane and so we want to compute the two-point correlator with both external legs on the UV brane. Finding this correlator is equivalent to computing, in the GNUV coordinates, the metric perturbation on the UV brane as a response to a source localized on the UV brane too.

\hspace{-.2cm}To obtain the two-point function we closely follow the work of Garriga and Tanaka~\cite{GT}. The
approach is to determine, at a point on the UV brane,
the linearized gravitational field  created by a source on the same UV brane. This can be related in a simple way to the graviton two-point correlator with
external legs on the UV brane. The key observation is that it is convenient to first work
in the GNTT gauge where the bulk equations are very simple. In this gauge the equations in the
bulk reduce to
\begin{equation}
\left(
\Box +  \partial_z^2 - {3 \over z} \de_z
\right)
h_{\mu \nu}
=
0 ~.
\label{bulkTTeqn}
\end{equation}
However, in this gauge
both branes
will in general not be straight and the bending
of each brane provides an additional
contribution to the stress tensors on the two boundaries, modifying
the boundary conditions.
The main effort of
these sections is to determine
this modification. With that information
and the solution to the propagator in the GNTT gauge,
we can readily evaluate the perturbation
in the GNIR(GNUV) gauge .

We consider below two cases. In the
first example, both the Fierz--Pauli mass term
and the source are located on the UV brane, and the perturbation
on the UV brane is determined. In the second, the Fierz--Pauli
mass term is placed on the IR brane, with the source
still placed on the UV brane.

\subsection{Fierz--Pauli Mass term on the Planck Brane }

In GN coordinates around the Planck brane (GNUV),
the UV boundary condition is
\beq
\partial_z h^{GNUV}_{\mu \nu} {}_{|_{z=R}}
= \lambda_{UV} h^{GNUV}_{\mu \nu} {}_{|_{z=R}} - \kappa_5^2
\left(S_{\mu \nu}^{UV}- \frac{1}{  3} S^{UV}\, \eta_{\mu \nu}  \right) ~.
\label{UVbc}
\end{equation}
with
\begin{equation}
\lambda_{UV} =
\kappa_5^2 f_{UV}^4.
\end{equation}
In order to be able to solve the Einstein's equations in the bulk, it is
useful
to perform a
coordinate transformation in order to obtain a graviton perturbation
that is transverse and traceless.
The transformations (\ref{eq:redGGN5}) and (\ref{eq:redGGNmu}) to TT coordinates
yields a new boundary condition:
\beq
\partial_zh^{GNTT}_{\mu \nu} {}_{|_{z=R}}= \lambda_{UV} h^{GNTT}_{\mu \nu} {}_{|_{z=R}}
- \kappa_5^2  \Sigma_{\mu \nu}
\end{equation}
where the source term now includes a brane-bending contribution:
\beq
\Sigma_{\mu \nu}
=
S_{\mu \nu}- \frac{1}{3} \eta_{\mu \nu} S
-  \frac{2}{\kappa_5^{2}} \partial_\mu\partial_\nu \zeta
+{\lambda_{UV} \over \kappa_5^2}
 \left( R\, \partial_\mu\partial_\nu \zeta
+{2 \over R}\, \zeta  \eta_{\mu \nu}  -
\partial_\nu \zeta_{\mu} - \partial_\mu \zeta_{\nu} \right)~.
\end{equation}
The gauge parameters $\zeta$ and $\zeta^{\mu}$ are
chosen so that in the new frame the metric is TT.
Since $\Sigma_{\mu\nu}$ is the source for the metric perturbation in the GNTT gauge, it must
must be transverse and
traceless too. This leads to the two conditions below on the gauge parameters:
\begin{eqnarray}
        \label{eq:TT1}
(i) &
\lambda_{UV} \left( {6 \over R} \partial_\mu \zeta -
\partial_\mu \partial_\sigma \zeta^\sigma
+ \Box \zeta_\mu \right) = 0,
\\
        \label{eq:TT2}
(ii) &
-\frac{\kappa_5^2}{3}  S - 2 \Box \zeta  +
\lambda_{UV} \left( {8 \over R}  \zeta  + R   \Box \zeta - 2
\partial_\sigma \zeta^\sigma \right)
=0.
\end{eqnarray}
%

\subsubsection{Massless case: $\lambda_{UV}=0$}

To begin though, first suppose that no mass term
is present. Then we should recover the results of Garriga and
Tanaka~\cite{GT}.
In this case
the TT conditions simplify and reduce to the single requirement
that
$\Box \zeta =-\kappa_5^2 S/6$. In the GNTT coordinates, the source is
thus related to the
brane stress-energy stress-energy tensor by:
\beq
        \label{eq:StoZ}
\Sigma_{\mu \nu} = S_{\mu \nu} - \frac{1}{  3} \left(\eta_{\mu \nu}
- \frac{\partial _{\mu} \partial_{\nu}}{  \Box} \right) S
\end{equation}
which is manifestly transverse and traceless. The
solution for the metric fluctuations
in the bulk is:
\beq
h^{GNTT}_{\mu \nu}(x,z)
=
-\kappa_5^2 \int d^4 x^{\prime }
\Delta(x,x^{\prime};z, R) \Sigma_{\mu \nu}(x^{\prime})
\end{equation}
where $\Delta$ is the Green's function for a scalar field
in the Randall--Sundrum background. It satisfies
the boundary conditions $\partial _z \Delta_{|_{z=R}}= \partial _z
\Delta_{|_{z=R^{\prime}}}=0$ and it's
solution may be found in the appendices.
Back in the GN system the metric perturbation on the brane is
\beq
h^{GNUV}_{\mu \nu}(x,R) =
h^{GNTT}_{\mu \nu}(x,R) +
\eta_{\mu \nu} \frac{\kappa^2_5 }{  3 R} \frac{1}{ \Box}\, S
\, .
\end{equation}
where we have substituted for $\zeta$ and
dropped terms involving longitudinal four-dimensional derivatives.
At long distances (see the Appendix, eq.~(\ref{eq:RS1prop})), $q \ll R^{-1}$,
the propagator becomes
\beq
\Delta(x,R;x^{\prime},R)
\rightarrow {2 \over R} \,
{1 \over 1 -(R/R^{\prime})^2}
\frac{1}{\Box} \delta^{4}(x-x^{\prime})
\end{equation}
and the metric perturbation in the GNUV coordinate is then~\cite{GT} (again dropping terms involving longitudinal four-dimensional derivatives)
\beq
h^{GNUV}_{\mu \nu} = -{2 \kappa^2_5 \over R}
\frac{1}{  1 - (R/R^{\prime})^2}{1 \over \Box}
\left( S_{\mu \nu} - \frac{1}{  2} \eta_{\mu \nu} S \right) -
{\kappa^2_5  \over 3 R}
\left(\frac{ R}{  R^{\prime}} \right) ^2
{1 \over 1 - (R/R^{\prime})^2} \frac{\eta_{\mu \nu}}{  \Box} S
\,.
\label{rsi}
\end{equation}
The crucial factor of 1/3 from
the gauge transformation $\zeta$ has been combined with the
$1/3$ factor appearing in the trace part of the Green's function
to obtain the correct factor of
1/2 for a massless graviton \cite{GT}. The part that is left over
is interpreted as due to the exchange of the radion,
and appears here with the correct sign to describe
a physical propagating
particle.

The important point in this review of the results
of Garriga and Tanaka is
to draw attention to the technical reason for recovering
the correct tensor structure of the massless graviton:
the transformation between the GNTT and GNUV coordinate
system involved a bending of the brane, $\zeta$, was
non-vanishing. By contrast, a transformation involving $\zeta_{\mu}$
can only modify the part of the graviton
propagator involving derivatives of the
source, leaving the part involving the
trace untouched. This is the
situation encountered when, on the brane, a graviton mass term
is turned on.

\subsubsection{Massive case: $\lambda_{UV} \neq 0$}

For a non-zero mass term on the brane the
first requirement~(\ref{eq:TT1}) becomes non-trivial
and implies that
\begin{equation}
\Box \zeta  =0 ~.
\end{equation}
Decomposing the vector $\zeta_\mu$ into a scalar
and a transverse part,
$\zeta_{\mu} = \zeta^T_{\mu}  - \partial_{\mu} \phi$ with
$\partial^\sigma \zeta^T_\sigma=0$,
the second condition~(\ref{eq:TT2})
relates the scalar part to the brane stress-energy tensor:
\bea
\Box \phi & =&  \frac{\kappa_5^2}{6 \lambda_{UV}} S - {4 \over R}
 \zeta ~.
\eea
A consistent solution to the TT conditions
is to set $\zeta =\zeta^T _{\mu}=0$. This
leads to the same expression~(\ref{eq:StoZ})
for the source $\Sigma_{\mu\nu}$ in terms of the boundary
stress--tensor $S_{\mu\nu}$.
Crucially though,
the coordinate transformation needed to reach the GNTT frame
now involves $\zeta_\mu$ rather than $\zeta$.
Thus there is no brane bending to compensate the
5D structure of the brane propagator and we expect that
gravity is never Einsteinian on the brane.

Going back to the GNUV
frame using
\beq
h^{GNUV}_{\mu \nu}=h^{GNTT}_{\mu \nu} - 2 \partial_{\mu} \partial_{\nu} \phi,
\end{equation}
we get a different expression for the metric fluctuation
compared to when no brane mass term is present:
\beq
h^{\scriptscriptstyle GNUV}_{\mu \nu}(x,z) = - \kappa_5^2 \int d^4x'
\Delta(x,x^{\prime};z,R)
\left(
S_{\mu \nu} - \frac{1}{3} \left(\eta_{\mu \nu}
- {\partial _{\mu} \partial_{\nu} \over \Box} \right) S
\right)
- \frac{\kappa_5^2}{3 \lambda_{UV}}\frac{\partial_{\mu}
\partial_{\nu}}{\Box} S.
\label{UVprop}
\end{equation}
Here $\Delta$ is the Green's function for a
scalar field in the Randall--Sundrum background
with a mass term on the UV brane. It satisfies the
boundary conditions $\partial _z \Delta_{|_{z=R}}= \lambda_{UV} \Delta_{|_{z=R}}$
and $\partial _z \Delta_{|_{z=R^{\prime}}}=0$.
It's expression is given in the appendix.

As already mentioned,
unlike the case with no brane mass term,
here there is no brane bending.
Thus the trace part of the propagator is the same
in the GNTT and GNUV coordinate systems; in particular,
the factor of 1/3 does not change and Einstein gravity
is not recovered.

In order to decouple the IR brane ($R/R^{\prime} \rightarrow 0$)
while keeping $f_{UV}$ held fixed,
we consider the long distance limit
$q R \ll 1$ but keeping $q R^{\prime} \gg 1$ in order to probe the fifth dimension.
Using the approximate expression~(\ref{eq:UVprop}) of the propagator found in
the Appendix for this limit, we arrive at
\beq
h^{GNUV}_{\mu \nu}(x,R) =- \frac{2 \kappa_5^2}{R}\,  \frac{1}{\Box -m^2}
\left(
S_{\mu \nu} -
\frac{1}{3} \left(\eta_{\mu \nu} - \frac{\partial _{\mu} \partial_{\nu}}{m^2}\right)S
\right),
\label{massRSII}
\end{equation}
where $m^2=2\lambda_{UV}/R$.
This is the correct propagator for a massive spin-2
particle \cite{boulwaredeser}, up to and including the derivative terms
that scale as $m^{-2}$. At distances $R\ll r \ll R^{\prime}$,
it is not surprising then to find
that the perturbation is dominated by the exchange of a single massive
spin--2, with the exchange of the KK tower suppressed as
in the Randall--Sundrum model.

At energy scales much below the compactification
scale, $r \gg R^{\prime}$, the theory is
four-dimen\-sional and the only light
states are the
radion and a massive
graviton, for which
there will be a vDVZ discontinuity. But in the limit just considered,
where the IR brane is decoupled first, we cannot appeal to these
arguments, as the theory is never
four--dimensional.
Nevertheless, the result above, (\ref{massRSII}),
demonstrates that even in this intrinsically five-dimensional
limit there is still a discontinuity.

Since in this limit there is a mass gap between the
would-be zero mode graviton and the
continuum of bulk gravitons that goes down
to zero, following the reasoning of  \cite{DRT} one may
suspect that the graviton studied here in
unstable. In the appendix the
two-point function is evaluated for complex, time-like
$q^2$. There we find that the light graviton
studied above does have a complex pole, with a
lifetime $\Gamma$ given by
$\Gamma/ m \sim (m R)^2$. This lifetime is parametrically identical
to the scalar example studied in \cite{DRT}.
Since here though $ R^{-1}  \gtap 10^{-3} $ eV and $m \ltap H_0$,
the lifetime is much longer than the age of the universe.

Finally,
one may be puzzled by the absence of any term in (\ref{UVprop})
that could possibly be interpreted as due to the exchange of
a radion with non-derivative couplings.
As shown at the end of Section 4, the
radion is normalizable and
physical (not a ghost), and has a wavefunction
that is localized about the IR brane. However, in
contrast to RS, here the radion wavefunction vanishes
at the UV brane and has only derivative couplings
to sources located there. Therefore it does not contribute
to the two-point correlation function of two conserved sources
located at the UV brane.

\subsection{Fierz--Pauli Mass term on the Infra-Red brane}

We consider the case where the Fierz--Pauli mass term is on the IR brane. The source remains
on the UV brane.

The first observation is that in the GGN coordinate system
with respect to the IR brane (GNIR), the metric
is additionally TT, since there is no source on the IR brane
(see Section 2).
Thus the metric satisfies (\ref{bulkTTeqn}) in the bulk, with
the IR boundary condition
\begin{equation}
\partial_z h^{GNTT}_{\mu \nu} {}_{|_{z=R^\prime}}
= - \lambda_{IR} {R \over R^{\prime}}\,
h^{GNTT}_{\mu \nu}{}_{|_{z=R^\prime}}  ,
\end{equation}
with
\begin{equation}
\lambda_{IR} =
\kappa_5^2 f_{IR}^4 .
\end{equation}

In this gauge however
the UV brane is bent, due to the source located there.
To determine the UV boundary condition in the GNTT gauge, we first
consider the GN coordinates with respect to the UV brane (GNUV) and then perform a coordinate transformation.
In the GNUV gauge, the UV boundary condition is
\begin{equation}
\partial_z h^{GNUV}_{\mu \nu} {}_{|_{z=R}} = - \kappa_5^2
\left(S_{\mu \nu} -
{1 \over 3} \eta_{\mu \nu} S \right).
\end{equation}
Inserting the
transformations (\ref{eq:redGGN5}) and
(\ref{eq:redGGNmu}) relating the GNUV gauge and GNTT gauge metric
into
the above boundary condition give
the desired UV boundary condition in the GNTT gauge:
\begin{equation}
\partial_z h^{GNTT}_{\mu \nu}{}_{|z=R} = -\kappa_5^2 \Sigma_{\mu \nu},
\end{equation}
where the source term now includes a brane-bending contribution:
\begin{equation}
\Sigma_{\mu \nu} = S_{\mu \nu} -{1 \over 3} \eta_{\mu \nu} S
-2 \kappa_5^{-2} \partial_{\mu} \partial _{\nu} \zeta ~.
\end{equation}
As before, requiring that this source is transverse and traceless fixes
$\Box \zeta = - \kappa_5^2 S/6$. In GNTT gauge then,
the UV source is
\begin{equation}
\Sigma_{\mu \nu} = S_{\mu \nu} -{1 \over 3} \left(
\eta_{\mu \nu}- {\partial _{\mu} \partial_{\nu} \over \Box} \right) S ~.
\end{equation}
The solution for the metric perturbation in the bulk is
\begin{equation}
h^{GNTT}_{\mu \nu}(x,z)
=
-\kappa_5^2 \int d^4 x^{\prime}
\Delta(x,x^{\prime};z,R) \,\Sigma_{\mu \nu}(x^{\prime}),
\end{equation}
where $\Delta$ satisfies the Green's function equation
\begin{equation}
\left(\Box + \partial^2_{z} - {3 \over z} \partial_z \right)
\Delta(x,z;x^{\prime},z^{\prime})={z^3 \over R^3}
\delta^{(4)}(x-x^{\prime})
\delta(z-z^{\prime}),
\end{equation}
with boundary conditions $\partial_z \Delta {}_{|z=R}=0$ and
$\partial_z \Delta{}_{|z=R^{\prime}}
+ \lambda_{IR} (R/R^{\prime} )\Delta {}_{|z=R^{\prime}}=0 $.
The Green's function $\Delta$ is given in the Appendices, and
more details can be found there.

Back in the GNUV coordinate system, the metric perturbation is
\begin{equation}
h^{GNUV}_{\mu \nu}(x,R)=h^{GNTT}_{\mu \nu}(x,R)+ \eta_{\mu \nu}
{\kappa_5^2 \over 3 R} {1 \over \Box}\, S,
\end{equation}
where we have substituted for $\zeta$ and dropped
terms involving four-dimensional derivatives. Using
results (\ref{eq:IRprop})-(\ref{eq:IRmass}) obtained in the Appendix, at long distances where we cannot probe the KK excitations,
$q R^{\prime} \ll 1$, the asymptotic form of the propagator is
\begin{equation}
\Delta(x,R;x^{\prime},R) \rightarrow   { \mathcal{N} \over \Box -m^2}\,
\delta^{(4)}(x-x^{\prime}),
\end{equation}
where $\mathcal{N}$ and $m^2$ may be found in the Appendix. Focusing on
non-derivative terms, in this limit the GNUV metric
perturbation reduces to
\begin{equation}
h^{GNUV}_{\mu \nu}(x,R)=- \kappa_5^2
  { \mathcal{N} \over \Box -m^2}
\left(S_{\mu \nu} - {1 \over 3} \eta_{\mu \nu} S \right)
+ \eta_{\mu \nu} {\kappa_5^2 \over 3 R} {1 \over \Box}\, S.
\end{equation}
The last term is due to the gauge transformation between
the GNTT and GNUV coordinates and is independent of the IR
boundary mass term.

In this limit the first two terms describe the exchange of
a massive graviton, and the last term describes the
exchange of a massless scalar (the radion).
But the sign of the last term implies that the radion is
ghost. This conclusion is independent of the size of the IR Fierz--Pauli mass, $f_{IR}$.

An interesting limit to look at is when  4D momenta can only probe the
lightest graviton and not the regular KK excitations:
$m  \ll q \ll  1/R^{\prime}$, then
$\mathcal{N} \rightarrow 2/(1-(R/R^{\prime})^2)/R$ and the metric perturbation on the UV brane becomes (again dropping terms involving longitundinal 4D derivatives)
\begin{equation}
h^{\scriptscriptstyle GNUV}_{\mu \nu}{}_{|R}
=
 -2\, {\kappa_5^2 \over R}
{1 \over 1- (R/R^{\prime})^2} {1 \over \Box}
\left( S_{\mu \nu} -\sfrac{1}{2} \eta_{\mu \nu}S \right)
-{\kappa_5^2  \over 3 R}\,
{\eta _{\mu \nu} \over 1- (R / R^{\prime})^2 }
{R^2 \over R^{\prime\, 2}}
{1 \over \Box }\, S,
\end{equation}
with ${\cal O}(\lambda_{{}_{IR}} R)$ corrections not included.
This expression goes smoothly over to the result for
Randall--Sundrum.

Finally, we can see that for UV brane
observers Einstein gravity is recovered at
distances shorter than the compactification scale.
Consider $qR^{\prime} \gg 1$ but $q R \ll 1$.
In this limit all dependence on the
IR brane disappears.
Using results~(\ref{eq:IRprop2}) obtained in Appendix, the
perturbation on the UV brane is indeed found to be (still dropping terms
involving longitundinal 4D derivatives)
\begin{equation}
h^{GNUV}_{\mu \nu}(x,R)=- {2 \kappa_5^2 \over R}
{1 \over \Box}
\left(S_{\mu \nu} - {1 \over 2} \eta_{\mu \nu} S \right).
\end{equation}
Nevertheless when $f_{IR} \neq 0$,
the theory
has a ghost which is responsible for the recovery of
4D gravity on the Planck brane. We will show in Section 4
that the ghost mode is the radion.

\section{Mode Decomposition Analysis}

\subsection{Spin-2 excitations: gravition mass spectrum}

We are first interested in the spectrum and the KK decomposition of the spin-2 excitations.
In the GNTT gauge, the bulk equations of motion do not couple the different polarizations and thus simply reduce to the a scalar equation of the form:
\beq
\label{sc1}
\Box \phi  +   \phi^{\prime\prime}  -  \sfrac{3}{z}  \phi^\prime  =  0.
\end{equation}
The mode decomposition can be written as
\beq
\phi(x,z) = \sum_{n} \,\left({z \over R}\right)^{2} \, \psi_n (z)  \phi_n (x),
\end{equation}
the wavefunctions $\psi_n (z)$ then satisfy a Bessel equation of order $\nu = 2$
($m_n$ is the 4D mass of the eigenmode):
\beq
\psi_n^{\prime \prime} + \sfrac{1}{z} \psi_n^\prime +  \left(m_n^2   \,
- \sfrac{4}{z^2} \right)
\psi_n   =  0,
\end{equation}
whose solutions are
\beq
\psi_m(z) =   A_n J_2(m_n z) + B_n  Y_2(m_n z),
\end{equation}
where the two constants are fixed by the boundary and the normalization conditions.
The boundary conditions for the spin-2 excitations are unaffected by coordinate transformations
of the form~(\ref{eq:redGGN5})-(\ref{eq:redGGNmu}) and therefore take the same form within
the GNTT gauge as in the GNUV and GNTT gauges:
\bea
	\label{eq:modeBCUV}
& \displaystyle
\left(
\psi_n^\prime +  \left( \frac{2}{R} - \lambda_{UV} \right)  \psi_n
\right)_{|z = R} = 0 ;
\\
	\label{eq:modeBCIR}
& \displaystyle
\left(
\psi_n^\prime  + \left( \frac{2}{R} + \lambda_{IR}  \right) \frac{R} {R^{\prime}}
\psi_n \right)_{|z = R^{\prime}}
 =  0.
\eea
Clearly as soon as a  nonvanishing mass term is turned on at either brane,  the would be massless mode, $\psi = R^2/z^2$, cannot satisfy the boundary conditions: the massless mode gets lifted by the brane localized masses. For the massive modes, the boundary
conditions~(\ref{eq:modeBCUV})-(\ref{eq:modeBCIR}) lead to the quantization equation:
\beq
	\label{eq:masseq}
\frac{m_n J_1(m_n R) - \lambda_{UV} J_2 (m_n R)}{m_n Y_1(m_n R) - \lambda_{UV} Y_2 (m_n R)}
=
\frac{m_n J_1(m_n R^\prime) + \lambda_{IR} \sfrac{R}{R^\prime} J_2 (m_n R^\prime)}{m_n Y_1(m_n R^\prime) + \lambda_{IR} \sfrac{R}{R^\prime}Y_2 (m_n R^\prime)}.
\end{equation}
Let us examine the solutions of this quantization equation in the two special cases when
a single brane mass term is turned on at either the IR or the UV brane.

\mathversion{bold}
\subsubsection{ Mass term on the IR brane ($\lambda_{UV} = 0, \lambda_{IR} \neq 0$)}
\mathversion{normal}

Assuming that $m_nR^\prime \ll 1$ and expanding the Bessel functions
near the origin, we find that the lightest mode has a mass approximately given by
\beq
\label{masscomp1}
m^2_0 = \frac{8}{R^{\prime\,2}}
\frac{\lambda_{IR} R}{ 4 + \lambda_{IR}R  } \left(\frac{R}{R^{\prime}} \right)^2.
\end{equation}
There is a gap of order $R/R^\prime$ between this lowest mode and the regular KK modes
that have mass
\beq
m_n \sim \frac{x_n}{R^\prime}
\end{equation}
where $x_n\sim (n+1/4)\pi$ are the roots of the $J_1$ Bessel function: $J_1 (x_n)=0$.

The normalization of the wavefunction can be found analytically using the Wronskian method
(see for instance Ref.~\cite{tit}). We found
\beq
\psi_n(z) \equiv  \psi_{m_n} = \mathcal{N}_n \,
\left(
J_2(m_n \, z) - \frac{J_1(m_n R)}{Y_1(m_n R)} \, Y_2(m_n \, z)
\right),
\end{equation}
with
\beq
\frac{1}{ \mathcal{N}_n^2} = \frac{2 }{ \pi^2 m_n^2 R}
\left(
\frac{ \lambda_{IR}^2 R^2 + 4 \lambda_{IR} R + m_n^2 R^{\prime\,2}}{
(\lambda_{IR}R\, Y_2(mR^\prime)+ m_n R^\prime\, Y_1 (m R^\prime))^2}
- \frac{1}{Y^2_1(m_n R)}
\right).
\end{equation}
%

\mathversion{bold}
\subsubsection{ Mass on the UV brane  ($\lambda_{IR} = 0, \lambda_{UV} \neq 0$)}
\mathversion{normal}

For a large warp factor, $R^\prime/R \gg1$, the quantization equation~(\ref{eq:masseq}) can approximately simplified to
\begin{equation}
(m_n Y_1(m_n R) - \lambda_{UV} Y_2 (m_n R)) \, J_1(m_n R^\prime)
= 0,
\end{equation}
the solutions of which form the regular KK modes again obtained from the roots, $x_n$, of the $J_1$ Bessel function:
\beq
m_n \sim \frac{x_n}{R^\prime}.
\end{equation}
On top of this tower, there is another mode that is continuously connected to the massless graviton when
$\lambda_{UV}$ goes to zero. For this special mode to be parametrically lighter than the
regular KK modes, the mass term added on the UV brane must be
small enough. More precisely, when $\lambda_{UV} R \ll (R/R^\prime)^2$, then the mass of the lightest graviton is approximated by
\beq
m_0^2 =  \lambda_{UV}R \left( \frac{R^\prime}{R}\right)^2 \frac{2}{R^{\prime\,2}}.
\end{equation}

The mass of this  mode can become
larger than the compactification
scale when $1/R^\prime$ gets smaller and smaller
and $\lambda_{UV} R $ held fixed.
In the limit $1/R^\prime \to 0$, it becomes non-normalizable and
is no longer in the spectrum.
Instead, a resonance with a finite lifetime is found (see Appendix A.3).

\subsection{The radion as a ghost}

\subsubsection{The radion wavefunction}

To provide further evidence that the interpretation
of the two--point
function obtained previously is indeed correct,
in this section the radion's wavefunction is determined and
its effective action computed. The principal
result of this section is a confirmation
that when the Fierz--Pauli mass term is
on the IR brane the radion is a ghost.

This conclusion is unchanged even if we allow 
for a non-Fierz-Pauli mass term on the 
IR brane. The details of that analysis are 
provided in Appendix B. 

In fact,
the wavefunction is rather straightforward to obtain
in the GNTT coordinate system. By Lorentz
covariance the metric describing  massless
scalar fluctuations $\Phi^i$ must be proportional
to $\partial_{\mu}\partial_\nu \Phi^i $ with $\Box \Phi^i=0$. Inspecting
Einstein's equations in GNTT coordinates the general
solution is trivial to obtain. It is :
\beq
h^{GNTT}_{\mu \nu}
=
-\frac{z^4}{2 R^3}\, \partial_{\mu}\partial_\nu f
+ \partial_{\mu}\partial_\nu \phi\, ,
\end{equation}
where $f$ and $\phi$ are massless scalars.
At this point the boundary conditions
are not yet imposed, since the branes are in general
bent. For future reference, in this
system the UV and IR branes are located at
$z_{UV}=R- \zeta(x)$ and $z_{IR}=R^{\prime}-(R^\prime/R)\, \phi_2(x) $
(the normalization is chosen for later convenience).

As a check on this result,
the wavefunction of Charmousis, Gregory and Rubakov (CGR)~\cite{CGR} for the
radion is now recovered.
To do this, transform to the GNUV coordinate
system $z^{(\scriptscriptstyle GNUV)}=z + (z/R)\, \zeta$
where the UV brane is straight and located
at $z_{UV}=R$. The IR brane is located at
$z_{IR}=R^{\prime}-(R^\prime/R)\,\eta(x)$
with $\eta=\phi_2-\zeta$.
The new metric is
\beq
h^{GNUV}_{\mu \nu} = - \frac{z^4}{2R^3} \, \partial_{\mu}\partial_\nu f
+ \partial_{\mu}\partial_\nu \phi
+ \frac{z^2}{R}\, \partial_{\mu} \partial_{\nu} \zeta
+{2 \over R}\, \eta_{\mu \nu }\,  \zeta.
\label{51}
\end{equation}
The UV boundary condition in this system is
$\partial _z h^{GNUV}_{\mu \nu}{}_{|z=R}=0$ and
determines the unknown function $\zeta$ to be
$\zeta =f$ with $\phi$ still unconstrained.
This gives the CGR solution
\beq
h^{GNUV}_{\mu \nu}=
\left(-\frac{z^4}{2R^3}  +  \frac{z^2}{R} \right) \partial_{\mu}\partial_\nu f
+ {2 \over R}\, \eta_{\mu \nu} \, f  + \partial_{\mu}\partial_\nu \phi\, .
\label{CGRradion}
\end{equation}
Actually this is not quite the CGR solution,
for here there is the additional
term proportional to $\phi$. In the RS1 model
with no brane mass term, this term
can be gauged away. This is because
in the restricted GN coordinate
system with the Planck brane straight and no
mass term on any brane, there is
a residual gauge invariance given by
$\zeta_{\mu}= \partial _{\mu} \phi/2$ that
may be used.

When the
IR brane mass term is present this gauge invariance
does not exist and $\phi$ cannot be eliminated in this
way. It is seen below
that in unitary gauge this mode is eliminated by the IR boundary
condition. In a non-unitary gauge $\phi$ corresponds
to the Goldstone boson associated to the
longitudinal component of the graviton.

To determine the radion function in the GNIR coordinate
system with the IR brane straight, it is
useful to recall the following result,
derived previously in Section 2. Namely, when the IR brane
mass term is present
the metric in these coordinates is
in addition TT.
But the most general
solution of Einstein's equations for a massless scalar
in GNTT gauge
was already
obtained. It is
\beq
h^{GNIR}_{\mu \nu}
=
h^{GNTT}_{\mu \nu} = \left(- \frac{z^4}{2R^3}  + \gamma \right)
\partial_{\mu}\partial_\nu f
+ \partial_{\mu}\partial_\nu \phi\, .
\label{radionwavefunction}
\end{equation}
(Compared with previous notation
here we have defined $\phi$ slightly differently and pulled out a factor $\gamma$).
This contains two massless scalars. One of these
is the radion and the other, as mentioned above, is the Goldstone boson
corresponding to the longitudinal component
of the graviton.
In the unitary gauge the
IR boundary condition is
$\partial_z h^{GNTT}_{\mu \nu} {}_{|_{z=R^{\prime}}}
= -\lambda_{{}_{IR}} (R^\prime/R) \, h^{GNTT}_{\mu \nu}{}_{|_{z=R^{\prime}}} $.
This determines $\phi $ in terms of $f$, or
equivalently, determines $\gamma$ after setting $\phi=0$.

In a non-unitary gauge $\phi$ is no longer zero.
But the IR boundary condition is then modified due to
an extra term coming from the Goldstone boson,
which may be identified with $\phi$.

In summary,
in GNIR gauge
(\ref{radionwavefunction}), with $\phi=0$, is the radion wavefunction.

\subsubsection{The radion kinetic term}

The wavefunction obtained above is not very
useful for determining the effective action,
since the UV brane is not straight. We would like
to compute the effective action in a coordinate
system with both branes straight.

To do this, begin in the GNIR gauge.
The radion wavefunction
is given by (\ref{radionwavefunction}), and the UV brane is
located at $z=R -\zeta(x)$.
We first need
to find the position of the
UV brane in the GNIR gauge. But this has been
determined already, since here the UV boundary condition
is the same as in RS1. Hence equation (\ref{51}) and (\ref{CGRradion}) yield
$\zeta =f$.

Next we straighten both branes. To do this,
it is easiest to start again in the GNIR=GNTT coordinates
where the IR brane is straight and the
radion wavefunction is given by (\ref{radionwavefunction}),
and perform a final coordinate transformation of the form
\beq
z^{(rad)} = z  - \frac{z\, F(z)}{R} \, f(x)
\label{finaltrans}
\end{equation}
maintaining $h_{\mu 5}=0$ but not $h_{55}=0$. $F$ is
an arbitrary function with the
only restriction that both branes
are now straight, which
implies $F(R^{\prime})=0$ and
$F(R)=-1$.
The normalization of
the radion wavefunction will be found
to depend only on the
values of $F$ at the location of the branes,
and not on its particular shape.
The metric in this final coordinate system is
\bea
& \displaystyle
h^{(rad)}_{\mu \nu} = c(z) \partial_{\mu}\partial_\nu f  -{2 \over R}  F(z) f \eta_{\mu \nu}\,,
\nonumber \\
& \displaystyle
h^{(rad)}_{55} = \frac{2z}{R} F^{\prime} f\,,
\label{radwave}
\eea
with
\beq
c(z)=- \frac{z^4}{2 R^3}  + \gamma -2 \int^z dz^{\prime} \, \frac{z^\prime}{R} F(z^{\prime})\, .
\end{equation}
It is straightforward to verify that for arbitrary
$F$, restricted to the boundary conditions $F(R^{\prime})=0$ and
$F(R)=-1$,  this expression satisfies the
equations in the bulk and also both boundary conditions.

An inspection of this solution indicates a significant
difference between the radion here and in the RSI model.
Here the brane mass term forces the non-TT part
of the radion wavefunction to vanish at the IR brane.
That is, here $F(R^{\prime})=0$. This is equivalent to
the requirement that the metric be traceless
in the GNIR coordinate system.
This fact is instrumental in turning the radion into
a ghost.

To determine the radion kinetic term we want to
integrate out the short-distance
variation of the metric. To do this we
follow the methodology of
\cite{PRZ},\cite{CF}. For full details, such as 
carefully adding the Gibbons--Hawking boundary term 
and seeing that the massive gravitons decouple, or 
for the more general case of a non-Fierz--Pauli mass term,   
see Appendix B. Here we quote the main results for  
the case of the Fierz--Pauli mass term. 
Expanding the action to quadratic
order gives
\beq
S_{eff} = - {1 \over 4 \kappa_5^2} \int d^4 x dz\, \left({R \over z}\right)^3 
h^{AB} {E_{AB}}[h_{CD}] +\hbox{boundary terms} +\hbox{brane mass term}~.
\end{equation}
Next we
insert the expression~(\ref{radwave}) for the radion
into the action, without using its four-dimensional
equations of motion. Since non-derivative
terms in the wavefunction satisfy
Einstein's equation, we are guaranteed that the
integrand is of the form $f \Box f$. It is
then a matter of collecting terms appearing
in the linearized Einstein equations
which have four-dimensional derivatives.
Then 
we find \footnote{We have added in the boundary terms. 
For more details see Appendix B.} 
\bea
S_{eff} &=& -{1 \over 4 \kappa_5^2}
\int d^4 x  \int ^{R^{\prime}}_{R} dz \, \left({R \over z}\right)^3 
 h^{AB} E_{AB}[h_{CD}] +\hbox{brane mass term} \nonumber \\ 
& & -{1 \over 8 \kappa_5^2} \int d^4 x \left({R \over z}\right)^3 
[h^{\mu \nu} h^{\prime}_{\mu \nu}- h h^{\prime} ]|^{R^{\prime}}_R
+{3  \over 8 \kappa_5^2} \int d^4 x {R^3 \over z^4} h_{55} h |^{R^{\prime}}_R \nonumber \\
&=&
-{3 \over 2 \kappa_5^2 R}\int d^4 x \,
f \Box f
\int ^{R^{\prime}}_{R} dz\,  F^{\prime} \nonumber \\
&=& -{3  \over 2 \kappa_5^2 R} \left( F(R^{\prime}) - F(R) \right)
\int d^4 x f \Box f
\label{finalkinterm}
\eea
Since $F(R^{\prime})=0$ and $F(R)=-1$, the radion kinetic
term is
\beq
S_{eff}=-{3  \over 2 \kappa_5^2 R} \int d^4 x f \Box f
\end{equation}
which has the wrong sign. The radion is a ghost!

Repeating this calculation for RS1
provides an independent check on the
overall sign, since here the radion is known to be
healthy. In fact, the formula is the
same and the boundary condition in the
UV is the same,  but the IR boundary condition is different.
So $F_{RS1}(R)=-1$ and
$F_{RS1}(R^{\prime})=-(R^{\prime\,2}/R^2)$, implying that
the radion has a physical kinetic term.

It is straightforward to repeat this exercise
when the Fierz--Pauli mass term is on the UV brane.
The radion wavefunction in the GNUV coordinates
is still given by (\ref{radionwavefunction}), but here
the integration constant $\gamma$ is different in
order to satisfy the UV boundary condition. Just
as in the previous example,
here it is the IR boundary condition that determines
the position of the IR brane in the GNUV gauge.
One finds
$\zeta =  (R^{\prime\,2}/R^2) f$, which not surprisingly, is
the same as in RS1.
Then starting
from GNUV coordinates, we straighten the IR brane,
keeping the UV brane straight and maintaining $h_{\mu 5}=0$.
In the notation of (\ref{finaltrans}), this requires $F(R)=0$ and
$F(R^{\prime})=- (R^{\prime\,2}/R^2)$. The computation
of the radion kinetic term proceeds as before, and
one arrives at (\ref{finalkinterm}).
Here though one finds that
the radion has a healthy kinetic term and is not a
ghost. The radion wavefunction is also
peaked at the IR brane and in the limit that the IR brane is decoupled
the radion is not normalizable. All of these
properties of the radion are also found
to occur in the RS model. These results with
a UV mass term are not
surprising, since all we are doing here is adding a
small perturbation on the UV brane where in the
RS model the radion already
had an exponentially small support.

Finally, a minor puzzle raised in Section 3 is now
resolved. In the computation (\ref{UVprop})
of the perturbation
due to a source on the UV brane there was no term
that could be interpreted as due to the exchange of
a radion with non-derivative couplings.
The reason for this is that the non-derivative component
of the radion appearing in $h_{\mu \nu}$ -- see~(\ref{radwave}) --
vanishes on the UV brane since in this model $F(R)=0$.

These conclusions generalize to the case with a non-Fierz-Pauli
mass term. Here we summarize the results of Appendix B. If
the mass term is on the IR brane, the radion is still a ghost but
now there is an additional state that is decoupled and has a physical kinetic term.
If the mass is on the UV brane, the radion still has a physical
kinetic term, but now there is an additional state that
is a ghost. Both of
these results are not surprising from the perspective of
the AdS/CFT correspondance. 

\section{Conclusions}

We have investigated the physics of brane-localized 
mass terms for the graviton in warped
backgrounds. We have performed a linearized analysis of the 
graviton two point correlator as well as a
mode decomposition of the five dimensional theory.  We find that 
if the mass term is localized on the
UV brane, observers on that brane see physics similar to that of 
a massive graviton in four dimensions.
One important distinction, however, is 
that if the graviton mass is larger than the mass of the
lightest Klauza-Klein modes it can now decay off the brane into these states.

A Fierz-Pauli 
mass term for the graviton on the IR brane 
reproduces Einstein's gravity for observers localized on
the UV brane at length scales shorter than 
the inverse mass of the lightest Kaluza--Klein modes. At
length scales longer than this the spectrum consists of a 
massive graviton and a ghost. It is the
radion field which is the ghost. 

For a non-Fierz-Pauli mass term on the IR brane 
there is an additional, physical state in 
the theory. But the radion field is still a ghost.  For a non-Fierz-Pauli mass term on the UV brane the radion is
physical but now there is an additional state in the theory that is a
ghost.

It is of interest to consider whether there are simple modifications
of this theory that could evade this problem. In models of latticized gravity {\cite{AS1}}
the radion excitation is absent, but unitarity is still maintained up to scales larger than the
compactification scale. It is therefore conceivable that a latticized version of the model we have
considered could be a successful realization of a theory which modifies gravity at long distances.

\section*{Appendix}
\setcounter{section}{0}
\setcounter{equation}{0}
\renewcommand{\thesection}{\Alph{section}}
\renewcommand{\theequation}{A.\arabic{equation}}

\section{Scalar Propagator}

\subsection{General expression}

The scalar Green's function equation with mass
terms localized on the UV and IR branes and a source
at $z=z^{\prime}$ is solution of the bulk equation
\beq
\left(
\partial^2 _z
-{3 \over z}\, \partial_z
-q^2
\right) \Delta
= {z^3 \over R^3} \delta (z-z^{\prime}),
\end{equation}
supplemented by the two boundary conditions at the UV and IR branes:
\begin{equation}
\partial _z \Delta {}_{|R}= \lambda_{UV}\, \Delta{}_{|R}
\ \ \ \mathrm{and} \ \ \
\partial_z \Delta {}_{|R^{\prime}}= - \lambda_{IR}\, {R \over R^{\prime}}\,
\Delta{}_{|R^{\prime}}.
\end{equation}

The Green's function
solution to this differential equation is obtained by first solving
the homogeneous equation to the left
($z < z^{\prime}$) and to the
right ($ z> z^{\prime}$) of the source. This gives
two solutions $\Delta_<$ and $\Delta_>$, respectively,
each having two undetermined integration constants.
The boundary conditions at the UV and IR
branes
fixes the ratio of the integration constants in each region.
Matching these two solutions at $z=z^\prime$ requires
continuity of the solution,
\beq
{\Delta_{<}}_{|z=z^{\prime}}={\Delta_{>}}_{|z=z^{\prime}}
\end{equation}
and the source equation implies
\beq
\partial_z (\Delta_{>}-\Delta_{<})_{|z=z^{\prime}} = {z^{\prime\, 3} \over R^{3}} ~.
\end{equation}
The first condition determines the ratio of integration
constants between the left and right regions, and the second condition
fixes their overall normalization.
The unique
solution, for space-like $q^2$, is
\beq
\Delta(z,z^{\prime})= {(z z^{\prime})^{2} \over R^3}
{1 \over \alpha \delta - \beta \gamma}
\left( \alpha K_{2}(q z_>)- \beta I_{2}(q z_>) \right)
\left(\gamma K_{2}(q z_<)- \delta I_{2}(q z_<) \right)\, ,
\label{propa1}
\end{equation}
where $z_< = \mathrm{Min} (z,z^\prime)$ and $z_> = \mathrm{Max} (z,z^\prime)$ and with
\bea
\alpha &=&
I_{1}(q R^{\prime})
+{\lambda_{IR} R \over q R^{\prime}} I_{2}(q R^{\prime}) \,,
\nonumber \\
\beta &=&
- K_{1}(q R^{\prime})
+ {\lambda_{IR} R \over q R^{\prime}} K_{2}(q R^{\prime}) \,,
\nonumber \\
\gamma &=&
-I_{1}(qR)
+{\lambda_{UV} \over q} I_{2}(qR)\,,
\nonumber \\
\delta &=&
K_{1}(qR)
+{\lambda_{UV} \over q} K_{2}(qR)\,.
\label{propa2}
\end{eqnarray}

\mathversion{bold}
\subsection{ Mass on the IR brane  ($\lambda_{UV} = 0, \lambda_{IR} \neq 0$)}
\mathversion{normal}

In the long distance limit $qR^{\prime} \ll 1$, by expanding the Bessel functions
around the origin we get the leading
form of the propagator with both legs on the UV brane
\beq
	\label{eq:IRprop}
\Delta(R,R,q^2) \rightarrow
{\mathcal{N} \over \Box - m^2}\, \delta^{(4)}(x^{\prime}-x)  ~,
\end{equation}
with
\beq
	\label{eq:IRnorm}
\mathcal{N}
=
\frac{2}{R}\,
\frac{1+\sfrac{1}{4}\left( 1 - \frac{R^4}{R^{\prime\,4}}\right) \lambda_{{}_{IR}} R}{
1- \frac{R^2}{R^{\prime\,2}} + \sfrac{1}{4}\left( 1 - 2 \frac{R^2}{R^{\prime\,2}}\right) \lambda_{{}_{IR}} R},
\end{equation}
and
\beq
	\label{eq:IRmass}
m^2 =
{8 \lambda_{{}_{IR}} \over R(4+ \lambda_{{}_{IR}} R)} \frac{}{}
 \left( {R \over R^{\prime}} \right)^4.
\end{equation}
to leading order in $R/R^{\prime}$.
We recover the expression~(\ref{masscomp1}) for the lightest graviton found in Section~4.1.
As another check, note that in the limit
$\lambda_{IR} R \rightarrow 0$, the RS1 result
$\mathcal{N} \rightarrow 2/(1- (R/R^{\prime})^2)/R$, is recovered:
\beq
	\label{eq:RS1prop}
\Delta(R,R,q^2) \rightarrow
\frac{2}{R(1-\frac{R^2}{R^{\prime\,2}})}\,
{1 \over \Box }\, \delta^{(4)}(x^{\prime}-x)  ~,
\end{equation}

In the limit $\lambda_{IR} R \ll 1$, we obtain
$m^2=2 \lambda_{IR}(R/R^{\prime})^4/R$. Using
$\lambda_{IR} = \kappa_5^2 f^4_{IR}$, where $f^4_{IR} $ is
the coefficient of the Fierz--Pauli (bare) mass term, gives
 $m^2= 2 (f_{IR} R/R^{\prime})^4/M^{2}_{Pl}$, the same result obtained in
the low--energy effective theory in the mass insertion
approximation.
In the opposite limit,
$\lambda_{IR}R \gg 1$, $m^2= (R/R^{\prime})^4/R^2$, which is
independent of the brane mass term and is
always less than the compactification scale $1/R^{\prime}$.

At distances below the compactification length scale, $q R^{\prime} \gg 1$,
but still above the AdS length scale, $ qR \ll 1$, the
leading term in the propagator is
\beq
	\label{eq:IRprop2}
\Delta(R,R,q^2) \rightarrow \frac{2}{R} \, \frac{1}{\Box}\, \delta^{(4)}(x^{\prime}-x).
\end{equation}
%

\mathversion{bold}
\subsection{ Mass on the UV brane ($\lambda_{{}_{IR}} = 0, \lambda_{{}_{UV}} \neq 0$)}
\mathversion{normal}

Using the asymptotic properties of the Bessel functions, it
is straightforward to perform the long distance limit
$q R  \ll 1$ while still probing the extra dimension $q R^{\prime} \gg 1$,
and we obtain the asymptotic  form of the  propagator with both legs on the UV brane
\beq
	\label{eq:UVprop}
\Delta(R,R,q^2) \rightarrow
\frac{2}{R}\,  \frac{1}{\Box -  m^2}\, \delta^{(4)}(x^{\prime}-x) ~,
\end{equation}
where
\beq
m^2 =2 { \lambda_{UV} \over R}~,
\label{m7}
\end{equation}
to leading order in $\lambda_{UV} R$.

While the validity of  this result requires $\lambda_{UV} R \ll 1$,
it does not restrict the relative size between
the graviton mass $m$ and the compactification scale $1/R^{\prime}$.
Thus we can use these results in the limit
that the IR brane is decoupled, $1/R^{\prime} \rightarrow 0$.
In this limit
there is a mass gap, with a continuum of bulk graviton states
down to 0. Following \cite{DRT}, we expect the massive graviton~(\ref{m7}) to
be unstable.
To see this one has to
compute the propagator
for time-like momenta $p^2 =-q^2 <0$.

Sending the IR brane to infinity,
$R/R^{\prime} \rightarrow 0$, and imposing
that positive frequency
waves are ingoing at $z = \infty$ (or equivalently,
performing the analytic continuation
of the propagator in (\ref{propa1}) and (\ref{propa2})), gives
\begin{equation}
\Delta(z,z^{\prime})
=
{\left(z z^{\prime} \right)^{2} \over R^3}
{ H^{(1)}_{2}(q z_>) H^{(1)}_{2}(qz_<) \, \mathcal{B}(qz_>) \over
q H^{(1)}_{1}(qR) -\lambda_{UV} H^{(1)}_{2}(qR)  }
\end{equation}
with
\begin{equation}
\mathcal{B}(qz_>)
=
q J_{1}(qR)- \lambda_{UV} J_{2}(qR)
-\left(qH^{(1)}_{1}(qR)- \lambda_{UV} H^{(1)}_{2}(qR)
\right)
\frac{J_{2}(qz_<)}{H^{(1)}_{2}(qz_<)}
\, .
\end{equation}
$H^{(1)}_\nu= J_\nu + i Y_\nu$ is the Hankel function of the first kind of order $\nu$.

As a check, note that  in the limit of a vanishing Fierz--Pauli mass, $\lambda_{UV}=0$, we recover the RS2 propagator
found in~\cite{GKR}.

The interesting result is the presence of a pole at
\beq
q\, {H^{(1)}_{1}(qR) \over H^{(1)}_{2}(qR)} - \lambda_{UV}=0 ~.
\label{complexpole}
\end{equation}
This is almost identical to the
equation solved by \cite{DRT} in a related context.
There they found a complex pole.
Following \cite{DRT},
we expand
this equation in the $qR \ll 1$ limit
using asympotic properties of
the Bessel functions and
\beq
 {H^{(1)}_{1}(qR) \over H^{(1)}_{2}(qR)}
= {Y_{1}(qR) \over Y_{2}(qR)}
\left(1 - i {J_{1}(qR) \over Y_{1}(qR)} + \cdots \right)
\end{equation}
where the ellipses denoted terms suppressed by $qR$.
The solution to (\ref{complexpole}) is given by
\beq
m= m_0 - i \Gamma
\end{equation}
with
$m^2_0= 2 \lambda_{UV} /R $ and $\Gamma/m_0 =\pi (m_0 R)^2/8$.

\section{Non-Fierz--Pauli mass term on the IR brane}
\setcounter{equation}{0}
\renewcommand{\thesection}{\Beta{section}}
\renewcommand{\theequation}{B.\arabic{equation}}

This Appendix analyses the gravitational spectrum 
for the case of a generic non-Fierz-Pauli mass term 
for the graviton on the IR brane. 
 
The bulk action is 
\beq 
S_{bulk} =  \int d^5x \sqrt{g} \left( \frac{\mathcal{R}}{2 \kappa_5^2}+ \cdots \right) 
\end{equation} 
where the $\ldots$ includes  in particular the Gibbons--Hawking boundary terms. The action 
on the IR brane is taken to be 
\beq 
S_{IR} =
{1 \over 8 \kappa_5^2} \left({R \over R^{\prime}}\right)^3 
\int d^4 x \, (a h^{ 2}_{\mu \nu} - b h^{2})_{|z=R^\prime} ~.
\label{nonFPmassterm}
\end{equation}
The case $a=b=-\kappa_5^2 f_{IR}^4 R/R'$ gives the Fierz--Pauli mass term studied in section~2. 
Since this brane action is not coordinate invariant, 
we need to specify the coordinates in which the action has 
this form. We choose it to describe the so-called 
GNIR coordinates, where $h_{55}=h_{\mu 5}=0$ locally near 
the brane. 

From the equations of motion we obtain the
boundary condition at the IR brane to be 
\beq 
(\partial_z h^{GNIR } _{\mu \nu} - \eta_{\mu \nu} 
\partial_z h^{GNIR}) |_{z=R^{\prime}} 
=(a h^{GNIR} _{\mu \nu} - b \eta_{\mu \nu} h^{GNIR} )|_{z=R^{\prime}} ~.
\label{nonFPbc}
\end{equation}

As in four-dimensional massive gravity with a non-Fierz-Pauli mass 
term, here we expect the existence of an additional 
propagating scalar degree of freedom, corresponding to the trace of the metric. 

Indeed, solving the bulk equations of motion and 
the boundary conditions allows for a non-zero trace of the form
\beq 
h(x,z)= 
\Phi(x) +{1 \over 6}(R^{\prime 2}-z^2)\, \frac{b-a}{a}\, \Box \Phi(x) 
\end{equation} 
where again $h=h^\mu_\mu$ and $\Box=\partial^\mu \partial_\mu$
and $\Phi$ is a 4D scalar field.
The boundary condition~(\ref{nonFPbc}) then simply determines the mass of $\Phi$:
\beq
m^2_{\Phi} = {a \over R^{\prime}} {(a-4 b) \over (b -a)} ~.
\label{phimass}
\end{equation}

Next, we would like to determine whether this field $\Phi$ is 
a ghost, and whether the radion is still 
a ghost when the mass term is not of the Fierz--Pauli 
form.  To this end, we will need to compute the off-shell 4D effective action.

First note that on-shell and in GNIR coordinates the most general solution to 
the bulk equations of motion and the 
IR boundary condition is given by  
\beq 
h^{GNIR}_{\mu \nu}=H_{\mu \nu}(x,z) + 
\left(- \frac{z^4}{2R^3}  + \gamma \right) 
\partial_{\mu} \partial_{ \nu} f(x)
+ \lambda_1 (z)  \partial_{\mu} \partial_{ \nu} \Phi(x) + 
\lambda_2 \Phi(x) \, \eta_{\mu \nu}
\label{nonFPmetricexpansion}
\end{equation} 
where the function $\lambda_1$ and the two constants $\lambda_2$ and $\gamma$ are given by
\begin{eqnarray}
& \displaystyle
\gamma ={R^{\prime 4} \over 2 R^3} -{2 \over a}{R^{\prime 3} \over R^3},\\
& \displaystyle
\lambda_2 = \frac{a-b}{3a},\\
& \displaystyle
\lambda_1 (z) = \frac{4b-a -\sfrac{1}{2} (z^2-R'^2)(b-a) m_\Phi^2}{3a m_\Phi^2}~.
\end{eqnarray}
Satisfying the boundary conditions 
and equations of motion implies that ({\it i}) $H_{\mu \nu}$ is 
transverse and traceless, ({\it ii}) $f$ is massless and it is identified with the radion of the previous sections, and ({\it iii}) $\Phi$ is 
the additional degree of freedom identified above with 
mass given by (\ref{phimass}). 

To find the effective four-dimensional action for these 
states we need to provide an  off-shell decomposition of the 
metric fluctuation $h_{\mu\nu}$. The decomposition~(\ref{nonFPmetricexpansion})
is unique once $f$ is related to the brane bending of the UV brane in the GNIR coordinates and once $\Phi$ is defined as the trace of the metric fluctuation at the boundary : 
\beq 
h^{GNIR}  |_{z=R^{\prime}} \equiv \Phi ~.
\end{equation}
This provides for an off-shell definition of 
the trace of $H_{\mu \nu}$.
 
The effective action is most easily computed 
in the coordinate system where the branes 
are parallel and fixed at $z=R$ and $z=R^{\prime}$ 
(so-called `rad' coordinates). 
Thus in the action given below, the metric appearing 
there is in the `rad' coordinates. 
The metric in these coordinates is obtained 
by transforming from GNIR coordinates to a coordinate 
system with both branes parallel. This gives  
\beq 
h^{rad}_{\mu \nu} = h^{GNIR}_{\mu \nu} 
-{2 \over R} \eta_{\mu \nu} F(z) \xi(x) -2 \int ^z _R dz {z^{\prime} \over R} F(z^{\prime}) 
\partial_{\mu} \partial_{\nu} \xi(x)~~, ~~h_{55}= {2z \over R} 
F^{\prime} \xi(x) 
\label{FPrad} 
\end{equation}
where 
\beq 
\xi(x) = f(x) +{R \over 6} \, \frac{b-a}{a} \, \Phi(x) 
\end{equation}
is 
the transformation needed to straighten the UV brane.

The five-dimensional action is given by 
\bea
S&=&-{1 \over 4 \kappa_5^2} \int d^5x \left( \frac{R}{z} \right)^3 h^{AB} E_{AB}[h_{CD}] 
 \nonumber
\\
& & - {1 \over 8 \kappa_5^2} \int d^4 x \left({R \over z}\right)^3 
[h^{\mu \nu} h^{\prime}_{\mu \nu}- h h^{\prime} ]|^{R^{\prime}}_R
+{3  \over 8 \kappa_5^2} \int d^4 x {R^3 \over z^4} h_{55} h 
|^{R^{\prime}}_R ~.
 \label{bulkactionFP}
 \eea
To this must be added the non-FP brane mass term (\ref{nonFPmassterm}). 
Each of these terms require some explanation. The variation of 
the first term gives (\ref{lineareeq1})--(\ref{lineareeq3}), 
the linearized equations of the motion in 
the bulk .
The terms in the second line are the linear equivalent of the 
Gibbons--Hawking terms: variation of the term on the first line produces  
terms 
on the boundary that are cancelled by the 
variation of the terms appearing in the second line.
All boundary terms of the type $O(\delta h^{\prime}_{AB})$ 
are cancelled this way. Terms that don't cancel 
are of the form $\delta h^{\mu \nu} {\cal O}_{\mu \nu}$. 
Requiring that they vanish gives the   
boundary conditions in `rad' coordinates. 
Using (\ref{FPrad}), one finds they are equivalent 
to the GNIR boundary conditions 
(\ref{nonFPbc}) that were previously  
inferred from the equations of motion. 
   
As previously mentioned, 
to this action must be added the non-Fierz--Pauli 
brane action. The only important point to note is 
that it must be evaluated in {\em GNIR coordinates}. 
(We could evaluate it in `rad' coordinates, 
but that would involve a lengthy substitution of 
$h^{GNIR}$ in terms of $h^{rad}$ into the brane action.) 

After a lengthy computation, substituting  
(\ref{FPrad}) into the bulk action (\ref{bulkactionFP}), 
using (\ref{nonFPmetricexpansion}), 
and including the brane mass term action (\ref{nonFPmassterm}),  
gives (without of course 
using the four-dimensional equations of motion)
\bea 
S_{eff} 
&=& 
-{1 \over 4 \kappa_5^2} \int^{R^{\prime}} _R d^5z 
{R^3 \over z^3} H^{\mu \nu} E_{\mu \nu}[H_{\rho \sigma}]
+ {a \over 8 \kappa_5^2} {R^3 \over R'^3}  
\int d^4 x \, (H^2_{\mu \nu}-H^2)_{|z=R'}  
\nonumber\\ 
& & 
-{1 \over 8 \kappa_5^2} \int d^4x {R^3 \over z^3} 
[H^{\mu \nu} H^{\prime}_{\mu \nu}- H H^{\prime} ]|^{R^{\prime}}_R
-{3 \over 2 \kappa_5^2 R} \int d^4x f \Box f 
\nonumber \\ 
& &
+{(b-a)^2 R^3 \over 24 a^2 \kappa_5^2 R^{\prime 2}} \int d^4x \, \Phi \Box \Phi
+ {(b-a) (4b-a) R^3 \over 24 a \kappa_5^2 R^{\prime 3}} \int d^4 x \, \Phi^2~.
\label{eq:Seff}
\eea
The first two lines describe the action for the massive gravitons 
and their (linearized) Gibbons--Hawking terms. Note that for 
the massive gravitons their mass term has been 
written in the Fierz--Pauli form. This guarantees 
that for these states there are five on-shell degrees of 
freedom. 
The last term in the second line and all the terms in the last
line describe the quadratic action for the radion and and the $\Phi$ field. 

We briefly highlight 
many significant cancellations that occured before arriving 
at this result.  
First note that both the radion and the $\Phi$ field have 
decoupled from each other 
and from {\em all} the spin-2 gravitons. This reassures us 
that at the quadratic level 
(\ref{nonFPmetricexpansion}) correctly decouples all the fields 
from each other. Further, all quadratic terms involving more 
than two derivatives also canceled. 

From the action~(\ref{eq:Seff}) we find that $\Phi$ is 
not a ghost, in contrast to what occurs in purely 
four dimensional massive gravity with a non-FP mass 
term. This may not be surprising, since in the 
AdS/CFT correspondence the non-FP mass term on the 
IR brane does not correspond in the CFT to adding a 
non-FP mass term, but rather to breaking general coordinate 
invariance in the IR.~\footnote{We have explicitly checked that when a non-Fierz--Pauli mass term is added on the UV brane the scalar field $\Phi$ is now a ghost as it could have also been guessed from the AdS/CFT correspondence.}
From the action~(\ref{eq:Seff}), we 
read off that the mass of $\Phi$ is given by  
\beq 
m_\Phi^2 = {a \over R^{\prime}} {(a-4 b) \over (b -a)} ~,
\end{equation}
which agrees with the previous computation using the 5D
equations of motion. This provides a non-trivial 
consistency check that
the computation of the effective action is correct. 

We find that even for the more 
general non-Fierz--Pauli mass term 
the radion is still a ghost. The value of its 
kinetic term is independent of whether or not the brane mass term 
has the Fierz-Pauli form. 

These results generalise 
our conclusion that the radion is ghost 
when the mass term has the Fierz--Pauli form. 
That is, in a theory with a 
non-Fierz--Pauli mass term on the IR brane 
the radion is always a ghost.

\section*{Note Added}

While this work was being completed, a work appeared~\cite{ghost}
that proposes a long distance modification of gravity based a Lorentz violating theory.
The model makes use of a ghost that condenses.  A connection between the presence of ghosts and Lorentz violations has also recently been studied in~\cite{CJM}

\section*{Acknowledgements}

We would like to thank Markus Luty, Jihad Mourad and Mark Wise for useful discussions.
ZC and MG would like to thank the hospitality of Saclay.
MG and CG would like to thank the hospitality
of Lawrence Berkeley National Laboratory.
CG and LP thank  the Aspen Center for
Physics  for its hospitality while part of this work was completed.
The work of MG is supported by the U.S.
Department of Energy under contract number
DE-FG03-92-ER40701. CG and LP are supported
in part by the RTN European Program
HPRN-CT-2000-00148
and the ACI Jeunes Chercheurs 2068.


\end{document}